\documentclass[a4paper, amsfonts, amssymb, amsmath, reprint, showkeys, nofootinbib, twoside]{revtex4-1}
\usepackage[english]{babel}
\usepackage[utf8]{inputenc}
\usepackage[colorinlistoftodos, color=green!40, prependcaption]{todonotes}
\usepackage{amsthm}
\usepackage{mathtools}
\usepackage{physics}
\usepackage{xcolor}
\usepackage{graphicx}
\usepackage[left=23mm,right=13mm,top=35mm,columnsep=15pt]{geometry} 
\usepackage{adjustbox}
\usepackage{placeins}
\usepackage[T1]{fontenc}
\usepackage{lipsum}
\usepackage{csquotes}
\usepackage{setspace}
\usepackage{siunitx}
\usepackage[english]{babel}
\usepackage[utf8]{inputenc}
\usepackage[pdftex, pdftitle={Article}, pdfauthor={Author}]{hyperref}

\setcitestyle{super,compress}

\begin{document}

\title{ATRP enhances structural correlations in polymerization-induced phase separation}

\author{Alba Sicher,\textsuperscript{[a,b]} Richard Whitfield,\textsuperscript{[c]} Jan Ilavsky,\textsuperscript{[d]} Vinodkumar Saranathan,\textsuperscript{[e]}  Athina Anastasaki,\textsuperscript{[c]} Eric R. Dufresne*\textsuperscript{[a]}}

 \affiliation{\vspace{0.5cm}$^{a}$Laboratory for Soft and Living Materials, Department of Materials, ETH Z\"{u}rich, Vladimir-Prelog-Weg 5/10, 8093 Z\"{u}rich, Switzerland.}

 \affiliation{$^{b}$Laboratory for Biomimetic Membranes and Textiles, Empa, Swiss Federal Laboratories for Materials Science and Technology, Lerchenfeldstrasse 5, 9014 St. Gallen, Switzerland.}
 
 \affiliation{$^{c}$Laboratory of Polymeric Materials, Department of Materials, ETH Z\"{u}rich, Vladimir-Prelog-Weg 5/10, 8093 Z\"{u}rich, Switzerland.}
 
 \affiliation{$^{d}$X-ray Science Division, Argonne National Laboratory, 9700 South Cass Avenue, Argonne, Illinois 60439, United States.}
 
 \affiliation{$^{e}$Division of Sciences, School of Interwoven Arts and Sciences, Krea University, 5655, Central Expressway, Sri City, Andhra Pradesh 517646 India.}
  
 \affiliation{}

 \email[Email: ]{eric.dufresne@mat.ethz.ch}

\vspace{1.6cm}

\begin{abstract}
\vspace{0.2cm} Synthetic methods to control the structure of materials at sub-micron  scales are typically based on the self-assembly of structural building blocks with precise size and morphology.
On the other hand, many living systems can generate structure across a broad range of length scales in one step directly from macromolecules, using phase separation.
Here, we introduce and control structure at the nano- and microscales through polymerization in the solid state, which has the unusual capability of both triggering and arresting phase separation. 
In particular, we show that atom transfer radical polymerization (ATRP) enables control of nucleation, growth, and stabilization of phase-separated poly-methylmethacrylate (PMMA) domains in a solid polystyrene (PS) matrix.
ATRP yields durable nanostructures with low size dispersity and high degrees of structural correlations.
Furthermore, we demonstrate that the length scale of these materials is controlled by the synthesis parameters.
\end{abstract}

\keywords{ATRP, polymerization,	phase separation, nanostructures, 	biomimetic synthesis}

\maketitle

\begin{figure}[h! tb] 
\centering 
\includegraphics[width=8.4cm]{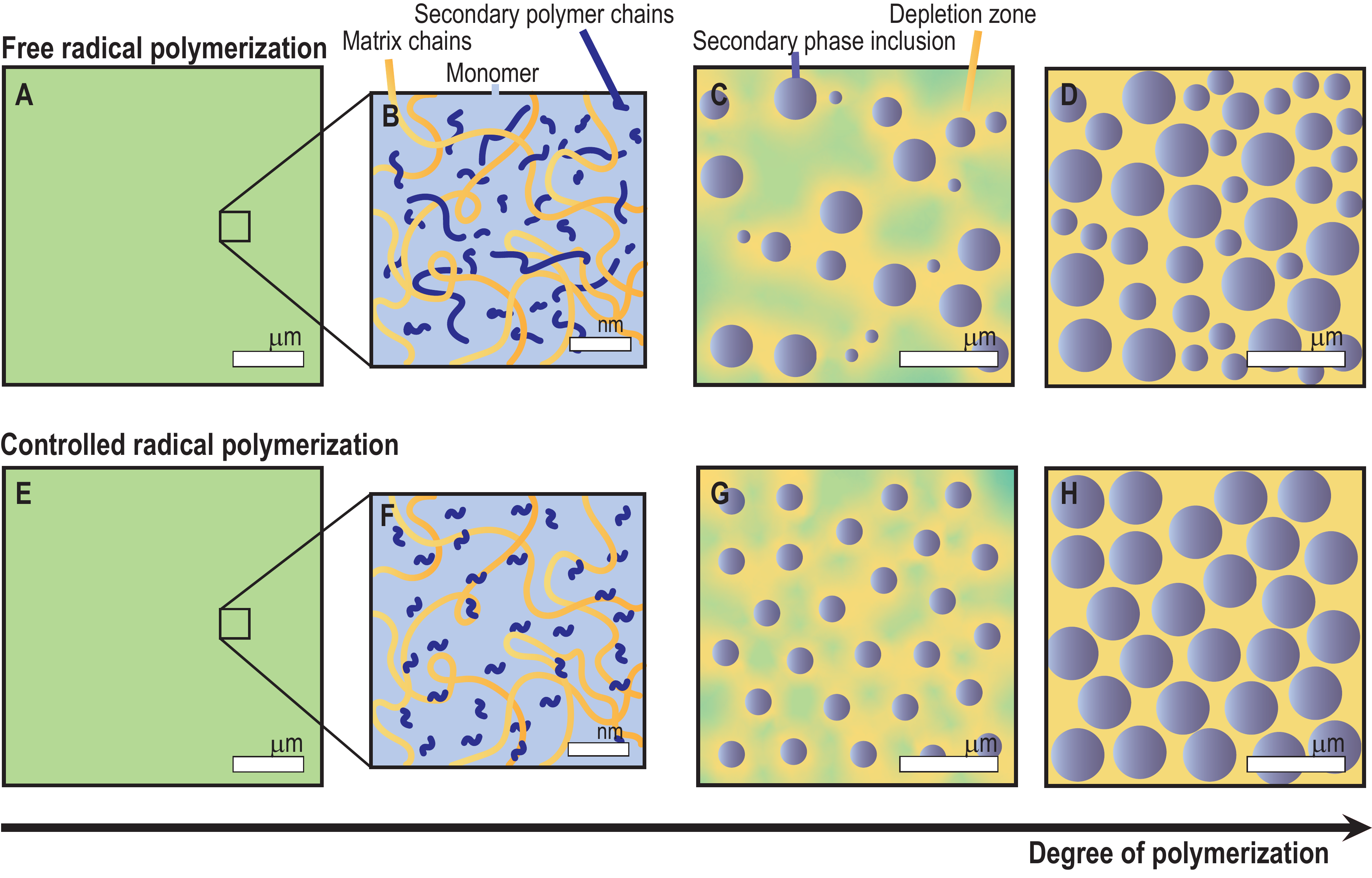}
\caption[Comparison of FRP- and ATRP-induced phase separation.]{\emph{Comparison of FRP- and ATRP-induced phase separation.} a,e) Uniform macroscopic conditions at the onset of the polymerization, before phase separation is triggered. A glassy matrix is swollen by a mixture of monomer and initiator. Green color indicates a mixture of blue and yellow components. b,f) Corresponding molecular views. We schematize the PS matrix in yellow, forming PMMA chains in blue, and MMA as a light-blue background pervading the system. c,g) Schematics of the microscopic phase-separated inclusions. The PMMA inclusions are blue, and the area around them where the monomer is depleted is yellow. d,h) Expected difference in the final structures and dispersity following kinetic arrest. 
}
\label{fig:atrp}
\end{figure}

\par Macromolecular materials often possess highly desirable properties that arise from their nano- and microstructure. 
For example, the filtration properties of membranes rely on their microscopic features, \cite{wienk1996recent, wang2013recent, haase2017multifunctional, fernandez2021putting} and the structural color of colloidal coatings depends on the size of the particles they contain. \cite{forster2010biomimetic, lee2017chameleon, shang2020photonic}
While synthetic nano- and micro-structured materials are commonly produced using self-assembly methods such as colloidal processing and block-copolymer assembly,  \cite{nagayama1996two, jiang1999single, matsubara2007thermally, forster2010biomimetic, lee2017chameleon, shang2020photonic, edrington2001polymer, valkama2004self, wu2018flexible, patel2020tunable} living systems generate structures in one step directly from macromolecules whose dimensions are much smaller than the characteristic scale of the final structure. 
For example, cells generate structures with pronounced order and low size dispersity through phase separation. 
Some examples include the photonic structures generating color in some bird feathers,   \cite{dufresne2009self, saranathan2012structure, prum2006anatomy, saranathan2021evolution} the topography of pollen's cell walls, \cite{radja2019pollen} and the regular porosity of diatom frustules. \cite{sumper2002phase, lenoci2008diatom, feofilova2022geometrical}

\begin{figure}[h! tb] 
\includegraphics[width=8cm]{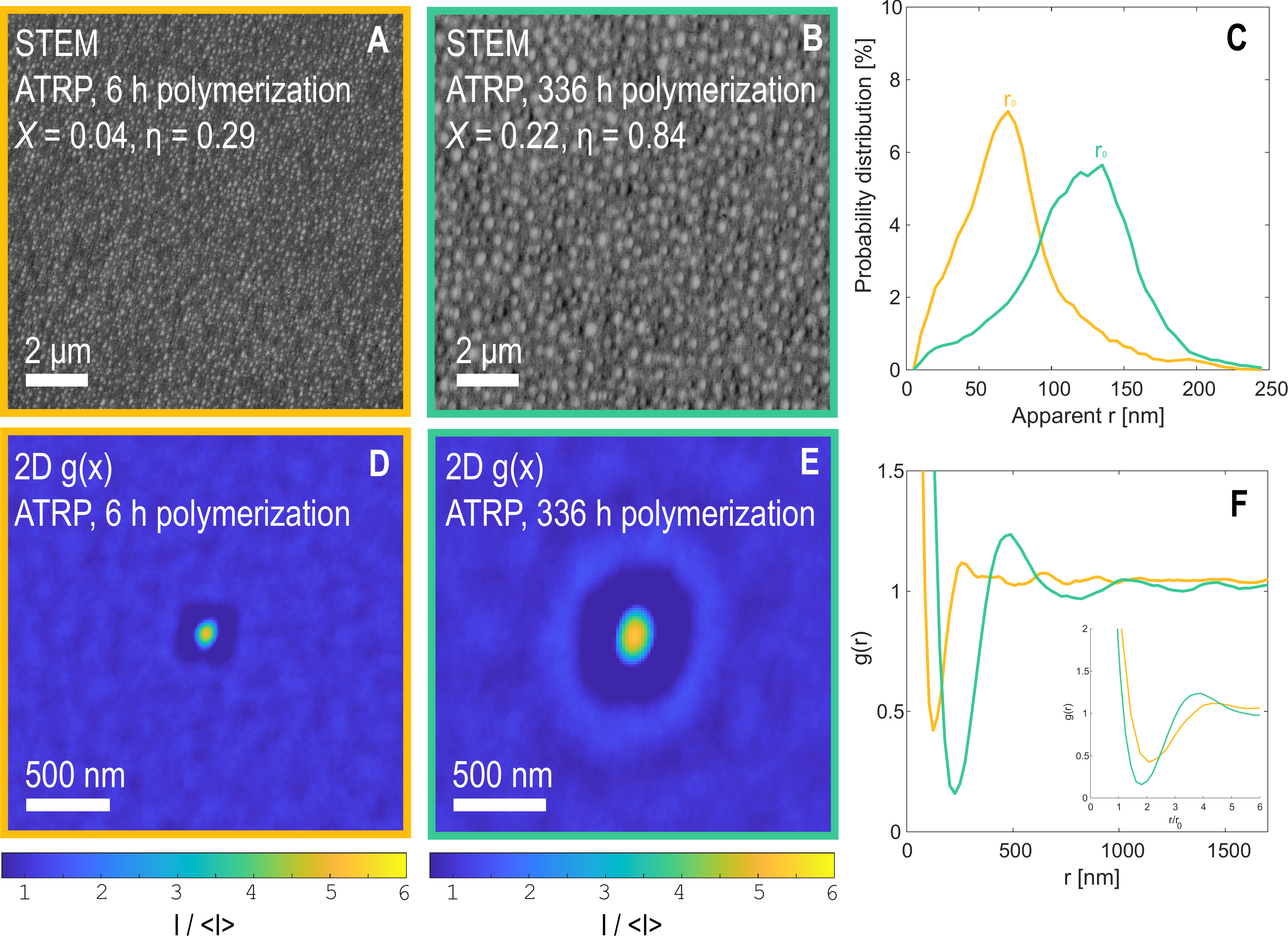}
\caption[Structure and correlations change during polymerization.]{\emph{Structure and correlations change during polymerization.} a, b) STEM images of two ATRP samples polymerized for different times. For each sample we report the polymerization time, the fraction of PMMA ($\chi$) and the degree of monomer conversion ($\eta$). c) Size distribution of the inclusions in panels a,b. On the x-axis: the apparent radius of particles on a 2D section of a 3D material. d,e) 2D distribution map calculated from the images in a and b, respectively. The colorbar is reported at the bottom of each panel. The rings in the distribution maps are distorted as a consequence of the fabrication of the PS film through hot-pressing. f) Azimuthal average of plots c,d. Inset: the x-axis of the azimuthal average is normalized by the mode of each particle size distribution (r$_0$). In panels c and f the yellow curves correspond to the sample in a,d and the green curves to the sample in b,e.}
\label{fig:timegr}
\end{figure}

We have previously shown that polymerization-induced phase separation in the solid state can be used to make durable, nanostructured, and colorful materials. \cite{sicher2021structural}
This process is inherently self-limiting, and can generate well-defined supramolecular structures in a single step.
Mechanistically, a monomer swells and plasticizes a polymeric glassy matrix, then polymerization induces phase separation and re-vitrifies the host matrix.
If the time scale of this transition is shorter than the one required for the complete demixing of the two polymer phases, phase separation is arrested in a kinetically trapped state.
With free-radical polymerization (FRP), the resulting composites appear blue or white in color, and have modest degrees of short-range  translational order.
This is attributed to FRP yielding broad molecular weight distributions through the continuous formation of new chains during the polymerization process. \cite{sheppard1979selection, matyjaszewski2012atom, bisht2001living} 
This constantly triggers phase separation as the chains become insoluble in the matrix, and results in phase-separated inclusions with broad size distributions (schematics in figure \ref{fig:atrp} a-d).

Here, atom transfer radical polymerization (ATRP) was used to provide control over the nucleation and growth of  phase-separated domains in the solid state.
Controlled- (or living-) radical polymerization, and in particular ATRP, relies on fast initiation and yields narrow molecular weight distributions through suppression of termination events, as well as slow and sustained growth. \cite{matyjaszewski2018advanced,goto2004kinetics, matyjaszewski2001atom,matyjaszewski2012atom, hill2015expanding, pan2016photomediated, rosen2009single}
In the solid state, we hypothesized that this would trigger the simultaneous nucleation of all the secondary-phase inclusions \cite{flory1942thermodynamics} (figure \ref{fig:atrp} f,g), yielding narrow size distributions.
We obtained durable nanostructures with significantly improved size dispersity and structural correlations, and we could tune the length scale in the material through control of the synthesis parameters.

\par To prepare the samples, we first hot-pressed 1 mm thick polystyrene (PS) films to form the glassy matrices.  We soaked them in mixtures of monomer (methyl methacrylate, MMA) and ethanol (EtOH) containing all the necessary ATRP components.
By diluting the reagents in ethanol, which does not partition into PS, we were able to tune their concentration in the matrix.
The used amount of MMA in EtOH was varied between 30\% and 40\% wt, and the total amount of solution was large compared to the quantity of monomer able to partition in the PS film. 
We selected activators regenerated electron transfer (ARGET) ATRP as our method, \cite{kwak2011arget, simakova2012aqueous, whitfield2021low} as this yielded a homogeneous polymerization, where high conversions, low dispersity and good control over molecular weight could be achieved with a ratio of [monomer]: [initiator]: [CuBr$_2$]: [ligand]: [Sn(2-ethylhexanoate)] = 200:1:0.01:0.01:0.1. \cite{chan2008arget}
Full details of the experimental procedure and reaction design are described in the supplementary information.

\par To understand how the structure develops, we used scanning transmission electron microscopy (STEM) and observed samples under indentical polymerization conditions at different time points. The results are visible in figure \ref{fig:timegr}.  Panel a shows the center of a sample at 6 hours.  At this time point,  polymerization and phase separation are ongoing. Panel b shows the structure at 336 hours, after polymerization has completed and the structure has reached steady state. The size distributions of PMMA domains are shown in panel c.   
As polymerization and phase separation proceed, the radii of the inclusions increase from $70~\mathrm{nm}$ to $130~\mathrm{nm}$.  

\par To quantify structural correlations of the PMMA inclusions, we calculated a two dimensional distribution map, $g(\vec x)$ \cite{mac2019application, boddeker2022non}.
We computed $g(\vec x)$ by averaging binarized areas of interest aligned with the center of each particle, and normalized with the average intensity in the binarized image, $\langle I \rangle$.  
The resulting image shows the average distribution of particle density around a fixed particle (figure \ref{fig:timegr} d, e).
 At early polymerization stages,  we observe a clear depletion zone around each particle (panel d).  
This makes sense, as  particles grow by depleting polymer chains from their surroundings. 
As polymerization proceeds, structural correlations emerge.
These correlations correspond to the bright ring around the center particle in panel e.  This ring indicates a regular center-center particle distance between the domains. In other words, it quantifies a characteristic structural length scale in the system.  
The azimuthal average of the distribution map yields a radial distribution function, $g(r)$, shown in panel f.
As the process proceeds, the peak of $g(r)$ moves to larger values of $r$.  This indicates that the domains not only grow, but their spacing increases from 260 to 490 nm.  Furthermore, the peak grows stronger as polymerization proceeds,  with the peak value of $g(r)$ increasing from 1.1 to 1.2. This shows that domains become more evenly distributed. 
Together, these features indicate that the particles retain some ability to rearrange within the matrix as they grow.

\par
\begin{figure}[tb] 
\includegraphics[width=8.4cm]{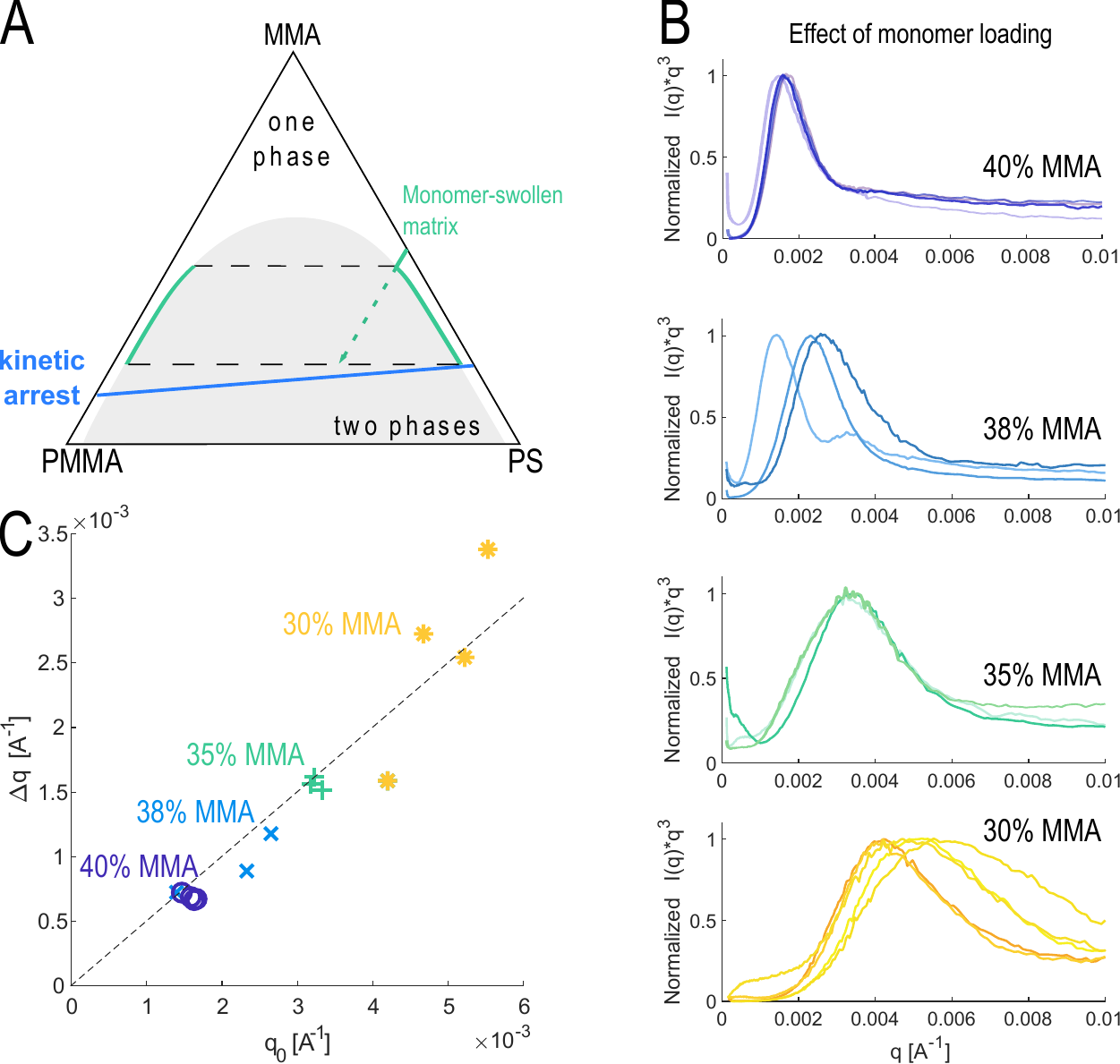}
\caption[Effect of monomer loading on the structure.] {\emph{Effect of monomer loading on the structure.} a) Hypothetical phase diagram. The two-phases region is colored in gray, the kinetic arrest line in blue, and the tie lines are dashed. The green path indicates the trajectory of a sample during polymerization. 
The dashed green line indicates the average composition, the solid green lines the composition of each phase. b) (U)SAXS profiles. Different lines on the same plot correspond to different samples synthesized in the same conditions at the same MMA-EtOH ratio. c) X-ray scattering peak width at 80\% height ($\Delta$q) plotted as a function of the peak position ($q_0$) for each curve in (b). The linear fit highlights scale-invariant samples with the same degree of structural correlations (same $\Delta q/q_0$).}
\label{fig:mmaetoh}
\end{figure}

  To gain some insight into the phase separation process,  we sketch the path that the system follows during polymerization on a hypothesized phase diagram (figure \ref{fig:mmaetoh} a).
This schematic diagram has a few essential features.  MMA and PS as well as  MMA and PMMA are fully miscible, while PS and PMMA are immiscible.  
Below a critical monomer concentration, the material undergoes kinetic arrest (\emph{i.e.} the matrix becomes glassy). 
The green path across the phase diagram shows the compositional shift of the two phases as the polymerization proceeds.
After swelling, the system starts off on the MMA/PS axis.
As polymerization starts, the system follows a contour of constant PS concentration. 
When the system reaches the boundary between the one- and two-phase regions of the phase diagram, it is saturated with polymer.  
With further polymerization, the system phase separates along the indicated tie-lines, creating PMMA- and PS-rich phases.
Ultimately, phase separation stops when the PS-rich phase reaches the kinetic arrest line and vitrifies.
In this picture, we expect the size of PMMA-rich domains to depend on the initial monomer loading. 
When relatively little monomer is loaded into the matrix, phase separation cannot proceed very far before the system vitrifies.

\par To quantify the effect of monomer loading on the microstructure, we used ultra-small-angle and small-angle X-ray scattering, (U)SAXS.
Results for MMA-EtOH ratios from 30 to  40\% are shown in figure \ref{fig:mmaetoh} b.
There, the scattering intensity, $I(q)$, is multiplied by the third power of the wavevector, $q^3$, to remove the characteristic large-$q$ decay of collapsed polymer coils. \cite{glatter1982small, saranathan2021evolution,perez2016structure}
For each condition, multiple samples were prepared, measured, and plotted individually.  
As expected, we found smaller length scales for low monomer loading.
Specifically, the position of the (U)SAXS peak shifted towards larger $q$ values as we reduced the amount of MMA in EtOH. 
The characteristic wavevectors  ranged from $0.0014$ to $0.0055$ \r{A}$^{-1}$.
In real space, these correspond to 450 and $110~\mathrm{nm}$, respectively.
The obtained structures are stable over a period of at least 9 months (figure S4).

\par As the scattering peaks shifted to larger $q$ values, they also broadened.
We make this trend more explicit by plotting the full width of the peak at 80\% height ($\Delta$q) as a function of the peak position ($q_0$) in figure \ref{fig:mmaetoh} c.
As the monomer loading is varied, the relative width of the scattering peak is constant. 
$\Delta q/q_0 \approx 0.5$ for all the samples, as indicated by the linear fit, although samples made with the lowest monomer loading (30\%) show larger sample-to-sample variability.
The consistency of $\Delta q/q_0$ suggests that the process of phase-separation and arrest is scale-invariant over the explored parameter range. \cite{saranathan2012structure,saranathan2021evolution}

\par
\begin{figure}[htb] 
\centering 
\includegraphics[width=8.4cm]{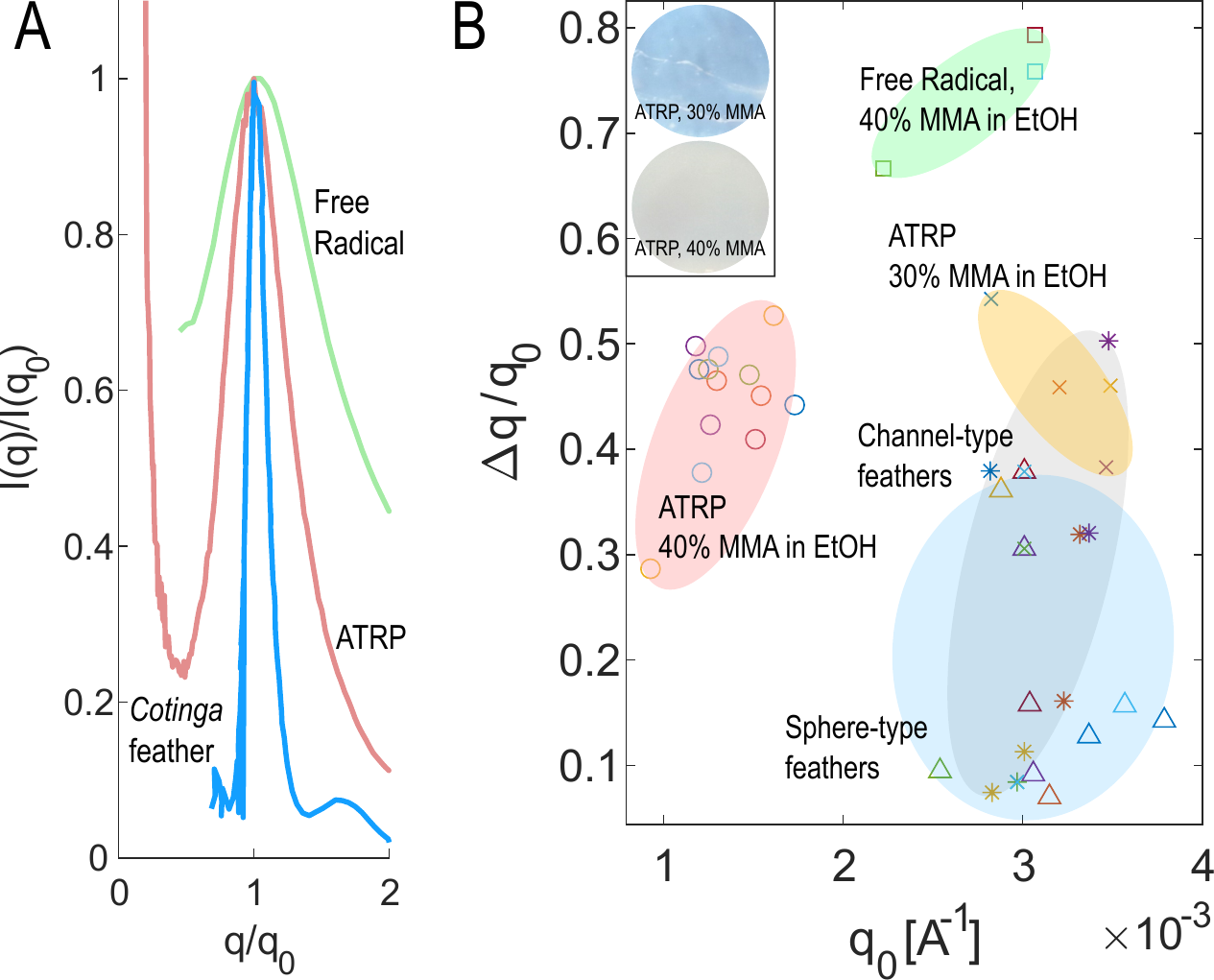}
\caption[Structural correlations in phase-separated materials.]{\emph{Structural correlations in phase-separated materials.} a) Normalized (U)SAXS curves corresponding to the nanostructure in the feather barbs of \emph{Cotinga maynana} (blue), and to two samples made with FRP (green) and ATRP (red), respectively. b) Diagram comparing the structural quality of PS-PMMA composites made with different strategies to biophotonic channel- (gray) and sphere-type nanostructures (blue) in bird feathers. Red: ATRP samples, 40\% MMA, 60\% EtOH. Yellow: ATRP samples, 30\% MMA, 70\% EtOH. Green: FRP samples. Inset: pictures of two ATRP samples made with 30 and 40\% MMA; 8 mm diameter. $\Delta$q is the full width at 80\% height. FRP data from \cite{sicher2021structural}.
}
\label{fig:atrpsaxs}
\end{figure}

 Next, we compare samples obtained with ATRP to those synthesized using FRP \cite{sicher2021structural} and to some biological nanostructures thought to form through polymerization-induced phase separation. \cite{saranathan2012structure} 
Figure \ref{fig:atrpsaxs} a shows the X-ray scattering intensities measured for a representative sphere-type bird feather (\emph{Cotinga maynana}), and for the best FRP sample and ATRP sample. While the scattering peak of the \emph{Cotinga} feather nanostructure remains unmatched, the ATRP sample shows comparable levels of structural correlations, with improved peak strength and narrowness compared to the FRP sample.

\par Figure \ref{fig:atrpsaxs} b shows $\Delta q/q_0$ as a function of peak position, and compares different classes of synthetic and natural materials.
Here, the peak width was measured directly from $I(q)$ instead of $q^3 I(q)$, because different materials have different characteristic decays at large $q$.
We find that $\Delta q/q_0$ is 0.74 and 0.42 for FRP and ATRP samples, respectively. 
The latter are comparable to channel-type biological samples, with the most strongly correlated ATRP sample ($\Delta q/q_0 \approx 0.3$) comparable to highly ordered sphere-type nanostructures. \cite{saranathan2012structure}
Additionally, ATRP samples made with 30\% MMA ($\Delta q/q_0 \approx 0.4-0.5$) have small enough inclusions to scatter blue light selectively, as shown in the inset.

\par In conclusion, we have designed a process that yields materials with spontaneously formed correlated structures at the sub-micron scale using phase separation.
Compared to FRP, ATRP offers tuneable size, and  stronger structural correlations. 
We expect that further reducing the nucleation time and molecular weight dispersity of polymeric chains will promote even lower particle size dispersity and higher level of order in the system.
More broadly, controlled-radical polymerization could become a powerful tool to manipulate phase separation for the creation of durable nanostructured materials at scale.

\section*{Experimental Section} 
\vspace{0.8em}

\textit{Preparation of the transparent PS samples}. Polystyrene pellets, average Mw $\sim 280,000$ Da (CAS: 9003-53-6) were purchased from Sigma-Aldrich. 
They were dried in vacuum at \SI{60}{\celsius} overnight in a Binder Vacuum drying oven, then hot-pressed into films of thickness 1 mm using a Fontune Holland table press. 
Pressing parameters: \SI{205}{\celsius} for 5 min, 40 kN. 
The films were then cut into pieces and annealed in vacuum at \SI{103}{\celsius} in the Binder oven to release stresses. \\

\vspace{0.8em}

\textit{Synthesis of the [Cu(TPMA)Br][Br] complex used in controlled radical polymerization.}
This procedure is based on the supplementary information of reference \cite{ribelli2018synthesis}.
200 mg of Cu(II)Br (Sigma Aldrich, CAS: 7789-45-9) and 260 mg of tris(2-pyridyl-methyl) amine (TPMA) (Sigma Aldrich, CAS: 16858-01-8) are dissolved in 200 ml of acetonitrile (Sigma Aldrich, CAS: 75-05-8).
The solution is stirred and degassed under nitrogen for 30 min, then sealed and stirred for an additional hour. 
Most of the liquid was then evaporated using a rotavapor (BUCHI, R-300) at \SI{45}{\celsius} and 140 mBar for 1 hour. 
The residual liquid was added drop by drop in diethyl ether (VWR, CAS: 60-29-7) and precipitated. 
The mixture was dried in vacuum over night. 
Afterwards, the dry [Cu(TPMA)Br][Br] complex was redispersed in a small amount of diethyl ether and stored in the freezer. 

\vspace{0.8em}

\textit{Fabrication of the PS-PMMA composites using ATRP}. 
Methyl methacrylate 99\% from Sigma-Aldrich (CAS: 80-62-6) containing mono-methyl ether hydroquinone as inhibitor was purified through filtration in a chromatographic column containing an inhibitor remover from Sigma-Aldrich (Product code: 311332). 
The polymerization procedure is based on paper \cite{chan2008arget}. 
The total molar ratio of the ARGET ATRP components is [Monomer]: [EBiB]: [CuBr2]: [TPMA]: [Sn(EH)2] = 200:1:0.01:0.01:0.1.
Three different solutions were prepared in Schlenk flasks: 1) A mixture of MMA, ethanol, butyl acetate (Sigma Aldrich, CAS: 123-86-4), and [Cu(TPMA)Br][Br]. The components where mixed with mass ratios 4:1:0.1:0.00025. 2) A mixture of ethanol and initiator, ethyl 2-bromo-isobutyrate (EBiB) (abcr, CAS: 600-00-0). The monomer- initiator ratio was 200:1. 3) A mixture of ethanol and tin(II) 2-ethylhexanoate Sn(EH)$_2$  (Sigma Aldrich, CAS: 301-10-0). 
In both solutions 2 and 3, the amount of ethanol was varied based on the total desired fraction of monomer and ethanol to be obtained in the final solution. 
The flasks were sealed with rubber septa and the solutions were degassed bubbling nitrogen through a Schlenk line. 
The vials where then sealed with vacuum and imported in a glove-box with a nitrogen environment. 
There, solutions 1 and 2 were mixed for about two minutes, before solution 3 was added to the mixture. 
The resulting solution was then split in different vials. 
The vials typically contained 20 ml of solutions, except for some of the samples prepared for figure 4 b, where different total amounts of solutions were used for the same MMA-EtOH ratio and ATRP reagents ratio.  
In the glove-box, the polystyrene films previously prepared, annealed, and left in the glove-box overnight to degas were soaked in the solutions. 
The vials were sealed and inserted into an oven in air, at \SI{70}{\celsius} for the desired polymerization time (typically 23 hours, unless otherwise specified). 
At the end of the polymerization, the samples were extracted from the swelling bath and deposited on glass petri dishes to allow the evaporation of any residual monomer. 

\vspace{0.8em}

\textit{STEM imaging}. Thin sections of 60 nm were obtained with a diamond knife (Diatome Ltd., Switzerland) on a Leica UC6 ultramicrotome (Leica Microsystems, Heerbrugg, Switzerland), and placed on Formvar and carbon coated TEM grids (Quantifoil, Großlöbichau, Germany). The sections were then coated with 6 nm of carbon using a Carbon Evaporator Safematic CCU-010. STEM analysis was performed using a ThermoFisher (FEI) Magellan 400 electron microscope. Electron micrographs of bird feathers were acquired according to reference \cite{prum2009development}. \\

\vspace{0.8em}

\textit{(U)SAXS experiments}. 
All (U)SAXS experiments on bird feathers were collected in transmission geometry at beamline 8-ID at the Advanced Photon Source as described in reference \cite{saranathan2012structure}.
SAXS measurements of the samples made using free-radical polymerization were performed at the cSAXS (X12SA) beamline at the Swiss Light Source (SLS, Paul Scherrer Institut) as described in \cite{sicher2021structural}. 
Absolutely-calibrated ultra-small angle X-ray scattering (USAXS) and small angle X-ray scattering (SAXS) experiments on samples made with controlled-radical polymerization were performed at the beamline 9-ID at the Advanced Photon Source, Argonne National Laboratory. 
The combined q range was $0.001-13~\mathrm{nm^{-1}}$, where $q = \frac{4\pi}{\lambda} sin(\theta)$. $\lambda$ is the wavelength and $\theta$ is half of the scattering angle. The X-ray energy was 21 keV ($\lambda = 0.5895$ \AA). X-ray photon flux was 5 x $10^{12}$ counts*mm$^{-2}$s$^{-1}$.\\

\vspace{0.8em}

\textit{NMR analysis}. Sample material was dissolved in deuterated chloroform >99.8\% from Apollo Scientific (CAS: 865-49-6). $^{1}$H-NMR was performed using a Bruker UltraShield 300 MHz magnet, and the results were analyzed using the software MestReNova. \\

\vspace{0.8em}

\textit{Synthesis of PMMA in bulk for size-exclusion chromatography (SEC)}.
A stock solution of CuBr$_2$ (2.09 mg) and TPMA (2.71 mg) was prepared in 1 mL of ethanol after 10 minutes of sonication. 
$100 \mu L$ (0.21 mg of CuBr$_2$ and 0.27 mg of TPMA, 0.01 equiv of CuBr$_2$/TPMA) of this stock solution, 2 mL of MMA (200 equiv.) and $13.7 \mu L$ (1 equiv.) of EBiB were added to a glass vial alongside a stirrer bar. 
This vial was sealed with a septum prior to bubbling with nitrogen for 20 minutes. 
In parallel, 30.3 mg of tin ethyl-2-hexanoate was dissolved in $250 \mu L$ of MMA. 
This was also sealed with a septum and degassed with nitrogen for 20 minutes. 
$27.5 \mu L$ of tin ethyl-2-hexanoate solution (3.03 mg, 0.1 equiv) was transferred into the main reaction mixture, via a nitrogen purged syringe. 
Polymerization was allowed to proceed at \SI{70}{\celsius} , with stirring set at 300 rpm. 
Samples were taken after 16 hours under a nitrogen blanket for 1H-NMR analysis and passed through a short column of neutral alumina to remove dissolved copper salts prior to SEC analysis. 
For all reactions, the same experimental procedure was followed, with various copper salts, ligands and initiators used. \\

\vspace{0.8em}

\textit{Size-exclusion chromatography (SEC)}. 
SEC analysis of PMMA samples synthesized in bulk was performed using a Shimadzu modular system comprising of a CBM-20A system controller, an SIL-20A automatic injector, a $10.0 \mu m$ bead size guard column (50 × 7.5 mm) followed by three KF-805L columns (300 × 8 mm, bead size: $10  \mu m$, pore size maximum: $5000$ \AA), an SPD-20A ultraviolet detector, and an RID-20A differential refractive-index detector. The temperature of the columns was maintained at \SI{40}{\celsius} using a CTO-20A oven. The eluent was N,N-dimethylacetamide (HPLC grade, with 0.03\% w/v LiBr) and the flow rate was kept at 1 ml*min$^{-1}$ using an LC-20AD pump. A molecular weight calibration curve was produced using commercial narrow molecular weight distribution poly(methyl methacrylate) standards with molecular weights ranging from 5000 to 1.5 × 106 g*mol$^{-1}$. Samples were filtered through 0.45 $\mu m$ filters prior to injection. 

\section*{Acknowledgements}

We thank Thomas Böddeker, Carla Fernandez-Rico, Nicolas Bain, Dominic Gerber and Yasir Mohammad for helpful conversations, Richard Prum, Alec Sandy and Suresh Narayanan for support with SAXS measurements of bird feathers, ScopeM (ETH Zurich) for access to electron microscopy, and the Laboratory of Soft Materials (ETH Zurich) for access to materials processing tools. This research used resources of the Advanced Photon Source, a U.S. Department of Energy (DOE) Office of Science user facility at Argonne National Laboratory and is based on research supported by the U.S. DOE Office of Science-Basic Energy Sciences, under Contract No. DE-AC02-06CH11357.

\section*{Conflict of Interest}
There are no conflicts of interest to declare.

\section*{Author contributions}
Each author contributed to this work as follows: A. S. and E. R. D. designed the project. 
A. S., R. W., and E. R. D. designed the polymer experiments. 
A. S., R. W. and J. I. performed and analyzed the polymer experiments. 
V. S. helped with the analysis of SAXS data and the comparison to biophotonic samples.
A. S., R. W., A. A and E. R. D. wrote the manuscript with input from all authors.


\section*{Supporting Information}
\renewcommand{\thefigure}{S\arabic{figure}}
\setcounter{figure}{0}

\section*{Reaction selection}

\begin{figure}[h!tb] 
\centering 
\includegraphics[width=8.4cm]{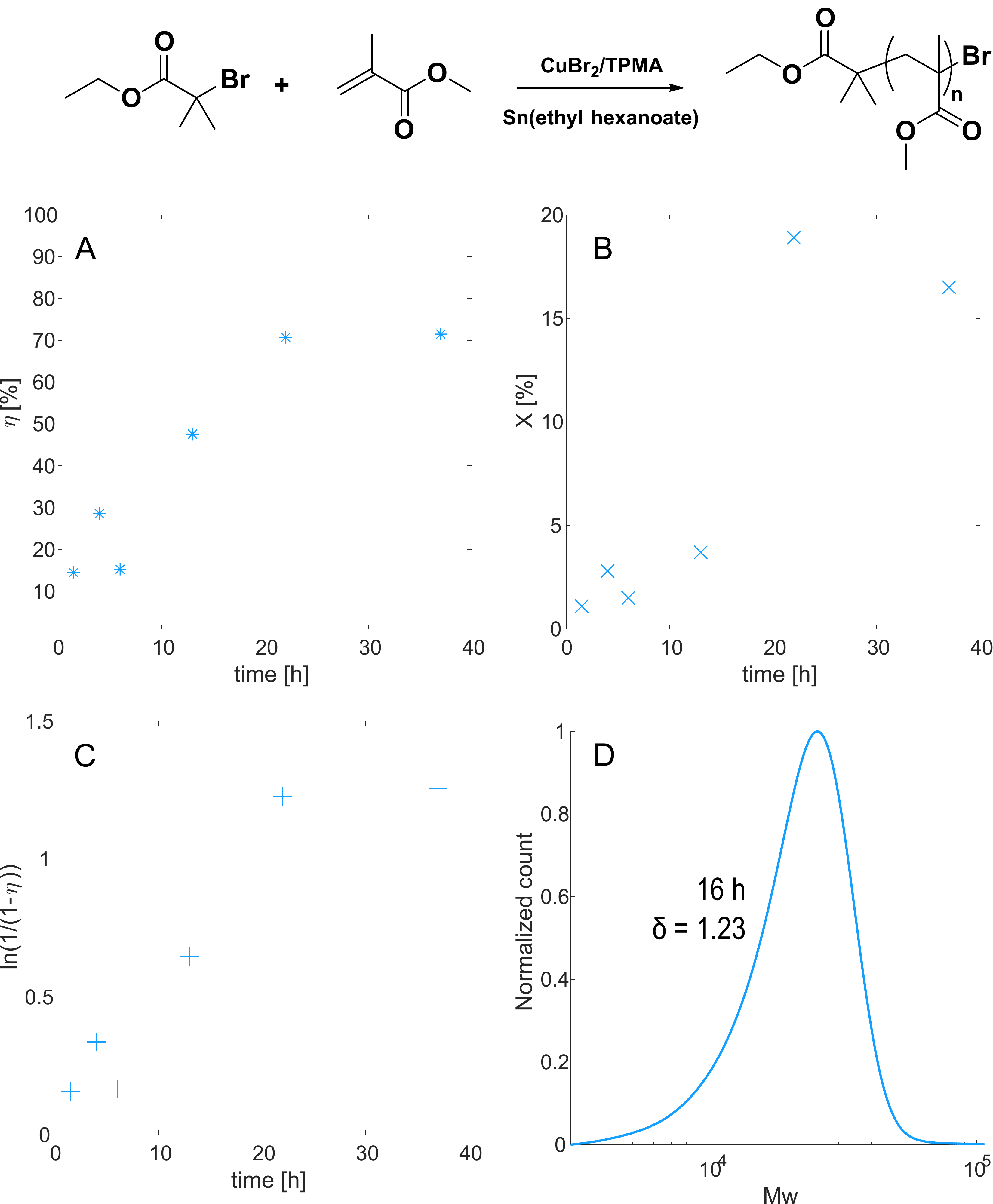}
\caption[Controlled-radical polymerization in the solid state.]{\emph{Controlled-radical polymerization in the solid state.} a) Conversion percentage of MMA to PMMA ($\eta$) over time during the polymerization. b) PMMA percentage in the PS-PMMA composite ($\chi$) during the polymerization. c) Natural logarithm of the initial amount of monomer over the current amount of monomer $(1-\eta)$. d) SEC curves measured at 16h into the polymerization. The molecular dispersity $\delta$ is reported. Reaction details: [MMA]: [EBiB]: [CuBr$_2$]: [TPMA]: [Sn(EH)$_2$] = 200:1:0.01:0.01:0.10, 
 \SI{70}{\celsius}, bulk. The schematics of the reaction is illustrated above the figure panels.}
\label{fig:SI_atrpgpc}
\end{figure}

\begin{table*}[h! tb]
	\caption{Selected polymerization reaction (figure \ref{fig:SI_atrpgpc}). \\ Reaction details: [MMA]: [EBiB]: [CuBr$_2$]: [TPMA]: [Sn(EH)$_2$] = 200:1:0.01:0.01:0.10, 
 \SI{70}{\celsius}, bulk \cite{chan2008arget}.}\label{tab3}
		\begin{tabular}{lccccccr}	
\hline		
Initiator & Ligand & Time (h) & Conversion (\%) & $M_{n(THEORY)}$& $M_{n(SEC)}$ & $M_{w(SEC)}$ & $\delta$ \\
\hline
EBiB 	&	 TPMA &	16	&	70	&	14200	&	18700 & 22900 &  1.23\\
\hline
	\end{tabular}
\end{table*}

\begin{table*}[h! tb]
	\caption{Reactions with different ligands and initiators (figure \ref{fig:SI_ligands}). \\ Reaction details: [MMA]:[Initiator]:[CuBr$_2$]:[Ligand]:[Sn(EH)$_2$] =200:1:0.01:0.01:0.10, \SI{70}{\celsius}, bulk  \cite{chan2008arget}.}\label{tab4}
		\begin{tabular}{lccccccr}	
\hline		
Initiator & Ligand & Time (h) & Conversion (\%) & $M_{n(THEORY)}$& $M_{n(SEC)}$& $M_{w(SEC)}$ & $\delta$ \\
\hline
EBiB 	&	 PMDETA &	16	&	60	&	12200	&	41000 & 801000 &  19.5\\
EBiB 	&	 HMTETA &	16	&	69	&	14000	&	52900 & 186000 &  3.52\\
MBPA 	&	 TPMA &	16	&	68	&	13800	&	12100 & 32500 &  2.69\\
\hline
	\end{tabular}
\end{table*}

\begin{table*}[h! tb]
	\caption{Reactions with a chloride initiator and catalyst (figure \ref{fig:SI_catalysis}). Reaction details: [MMA]:[Initiator]:[CuBr$_2$]:[Ligand]:[Sn(EH)$_2$] =200:1:0.1:0.15:0.05, \SI{70}{\celsius}, bulk \cite{martinez2021copper}.}\label{tab5}
		\begin{tabular}{lcccccr}	
\hline		
Ligand  & Time (h) & Conversion (\%) & $M_{n(THEORY)}$& $M_{n(SEC)}$& $M_{w(SEC)}$ & $\delta$ \\
\hline
PMDETA 	&	16 &	8	&	1800	&	1300 & 1500 &  1.15\\
PMDETA 	&	40 &	12	&	2600	&	1800 & 2200 &  1.19\\
TPMA 	&	16 &	16	&	3400	&	2300 & 3000 &  1.33\\
TPMA	&	40 &	29	&	6000	&	6400 & 7500 &  1.19\\	
\hline
	\end{tabular}
\end{table*}

\begin{figure}[h b] 
\centering 
\includegraphics[width=8.2cm]{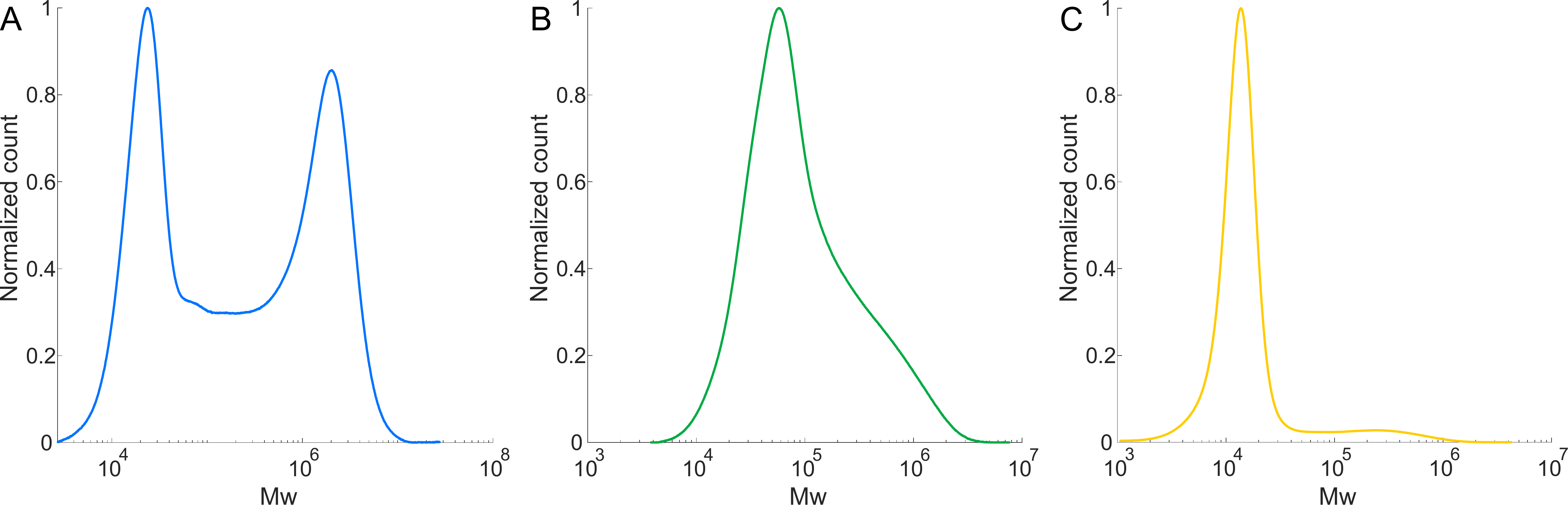}
\caption[Effect of the activity of ligand and initiator on the polymerization.]{\emph{Effect of the activity of ligand and initiator on the polymerization.} a) Molecular weight distribution corresponding to ligand PMDETA. b) Molecular weight distribution corresponding to ligand HMTETA. c) Molecular weight distribution corresponding to initiator MBPA. Reaction details: [MMA]:[Initiator]:[CuBr$_2$]:[Ligand]:[Sn(EH)$_2$] =200:1:0.01:0.01:0.10, \SI{70}{\celsius}, bulk.
}
\label{fig:SI_ligands}
\end{figure}

\begin{figure}[h tb] 
\centering 
\includegraphics[width=8cm]{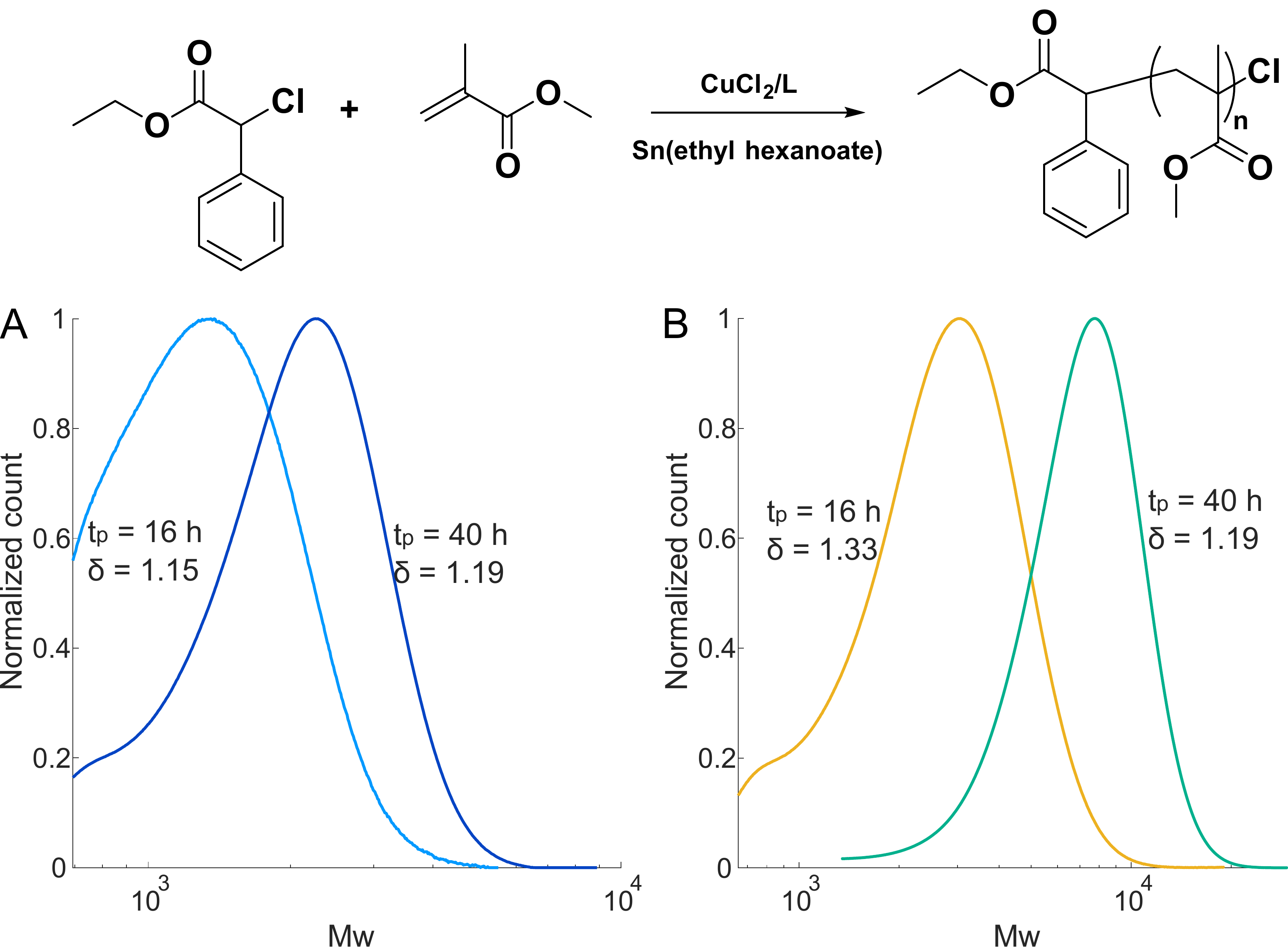}
\caption[Effect of the catalytic complex on the polymerization.]{\emph{Effect of the catalytic complex on the polymerization.} a) Molecular weight distribution corresponding to the complex CuCl$_2$/PMDETA. b) Molecular weight distribution corresponding to the complex CuCl$_2$/TPMA. For each curve, the polymerization time (t$_p$) and the molecular weght dispersity ($\delta$) are reported. Reaction details: [MMA]:[Initiator]:[CuBr$_2$]:[Ligand]:[Sn(EH)$_2$] =200:1:0.1:0.15:0.05, \SI{70}{\celsius}, bulk. The schematics of the reaction is illustrated above the figure panels. 
}
\label{fig:SI_catalysis}
\end{figure}

\par Here we provide supplementary information about the design of the polymerization reaction and the durability of the samples. We also describe in details the experimental procedures used.

\par We selected activators regenerated electron transfer (ARGET) ATRP as our polymerization method.
ARGET ATRP yielded a homogeneous polymerization where high conversions, low dispersity and good control over molecular weight could be achieved with a ratio of [MMA]: [EBiB]: [CuBr$_2$]: [TPMA]: [Sn(EH)$_2$] = 200:1:0.01:0.01:0.10 \cite{chan2008arget}.
We tested the reaction using NMR and size-exclusion chromatography (SEC). The results are schematized in table \ref{tab3}, and shown in figure \ref{fig:SI_atrpgpc}.
Panels (a) and (b) in figure \ref{fig:SI_atrpgpc} refer to polymerization in the solid state, and show the evolution over time of monomer conversion ($\eta$) and fraction of secondary polymer in the composite ($X$), expressed as percentages.
Most of the conversion and the phase separation occur in the first 24h of polymerization, while the fraction of PMMA in the composite grows to $X = 20\%$. 

\par Panel (c) of figure \ref{fig:SI_atrpgpc} shows that in our system $ln\frac{[1]}{[1-\eta]}$ is linear as a function of time for the first 24 hours of polymerization in the solid state. 
In fact, a marker of living radical polymerization is first-order kinetics, due to the identical rates of activation and deactivation of the polymerizing chains, and the negligible contribution of non-reversible termination \cite{matyjaszewski2001atom}. 
This kinetic behaviour is described by the following equation \cite{matyjaszewski2001atom}:

\begin{equation}\label{eq_atrp1}
\centering
R_p = -\frac{d[M]}{dt} = k_p[P][M] 
\end{equation}

\vspace{0.5em} 

where $R_p$ is the polymerization rate, defined as the variation of monomer concentration $[M]$ over time, $[P]$ is the concentration of the active propagating species, and $k_p$ is the propagation rate.
Therefore, assuming a constant number of chains, the monomer concentration decreases exponentially over time:

\begin{equation}\label{eq_atrp3}
\centering
[M] = [M_0] e^{-k_p[P] t} = [M_0] e^{-k_p^{app} t} \hspace{0.2cm} (if [P] = constant)
\end{equation}

\vspace{0.5em} 

where $[M_0]$ is the initial monomer concentration, and $k_p^{app}$ is the apparent propagation rate if [P] is constant, as expected for living polymerization.

To compare to experimental data, this is normally written as:

\begin{equation}\label{eq_atrp2}
\centering
ln\frac{[M_0]}{[M]} = ln\frac{[1]}{[1-\eta]} = k_p^{app}t 
\end{equation}

\vspace{0.5em} 

If [P] is constant, $ln\frac{[1]}{[1-\eta]}$ varies linearly with time, like in the first 24 hours of figure \ref{fig:SI_atrpgpc} c, whereas in a poorly-controlled reaction an upward or downward curvature in this kinetic plot indicate either slow initiation or termination and other side-reactions, respectively \cite{matyjaszewski2001atom}.

\par A second marker of controlled polymerization is a narrow molecular weight distribution. 
We thus performed SEC measurements, whose results are shown in panel (d). 
The curve corresponds to the molecular weight distribution of PMMA polymerized for 16 hours. 
This curve refers to the polymerization of pure monomer in bulk, not to polymerization in the solid state.
In fact, the presence of PS in the composite prevented the separation of PMMA and PS prior to and during SEC, where the curves for the two polymers are highly convoluted.
The molecular weight dispersity is 1.23, and further hints to the good control retained during the polymerization process \cite{parkatzidis2020recent}.
The dispersity is calculated as $\delta = M_w/M_n$, with $M_w$ being the weight averaged molecular weight, and $M_n$ the number average molecular weight.

\begin{figure}[h! tb] 
\centering 
\includegraphics[width=7cm]{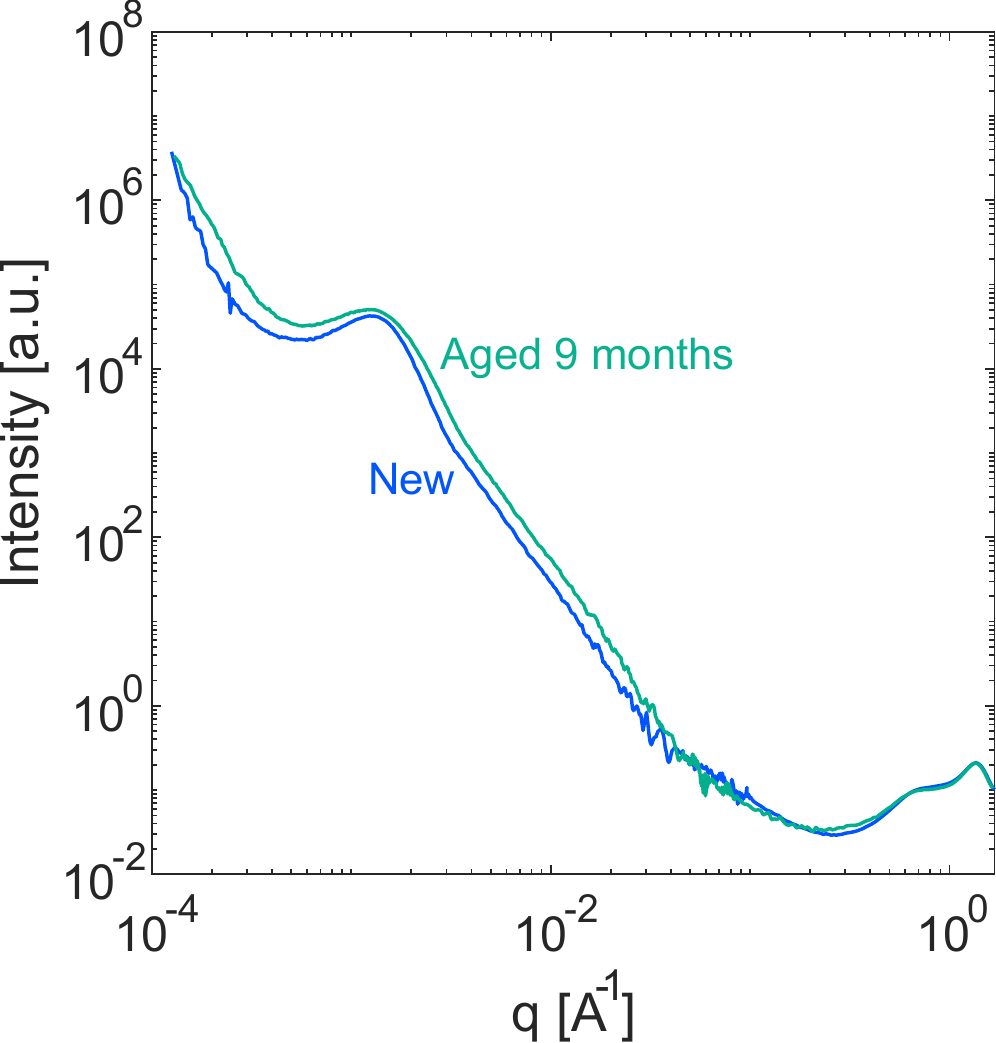}
\caption[Durability of an ATRP sample over time.]{\emph{Durability of an ATRP sample over time.}  (U)SAXS spectra of the same sample acquired right after the synthesis (blue) and after 9 months on a bench at room conditions (green). 
}
\label{fig:SI_age}
\end{figure}

\par The selected reaction worked well for our system compared to a broad range of screened reaction parameters. 
Conducting the polymerization with lower activity ligands, such as N\textquotesingle\textquotesingle,N\textquotesingle\textquotesingle-Pentamethyldiethylenetriamine (PMDETA) and 1,1,4,7,10,10-Hexamethyltriethylenetetramine (HMTE-TA) \cite{chan2008arget}, resulted in uncontrolled polymerization with the formation of large amounts of very high molecular weight polymer (figure \ref{fig:SI_ligands} a,b). 
Similarly, when EBiB was substituted for the higher activity initiator, methyl-$\alpha$-bromophenylace-tate (MBPA), a high molecular weight shoulder was observed in the SEC trace (figure \ref{fig:SI_ligands} c).  
Using CuCl$_2$/PMDETA and CuCl$_2$/TPMA as the catalytic complex resulted \cite{martinez2021copper} in controlled polymerization, with Mn values close to theoretical and low dispersity values (figure \ref{fig:SI_catalysis} a,b).
However, reactions were very slow, reaching less than 30\% conversion in 40 hours. Therefore these reaction conditions were not deemed suitable for our system, where high conversions are needed to arrest the phase separation process.

\section*{Durability of the structure} 

\vspace{0.8em}

In figure \ref{fig:SI_age} we show that samples are stable over time and show negligible ageing. 
We plot the X-ray scattering spectra of the same ATRP sample measured right after the systhesis and after 9 months. 
The peaks of the spectra collapse on top of each other, indicating the permanence of the length scale in the material over the probed period of time. 
The overall shape of the curves is also similar in the two cases, suggesting good structural stability across all length scales. 



\begin{thebibliography}{10}

\bibitem{wienk1996recent}
I.~Wienk, R.~Boom, M.~Beerlage, A.~Bulte, C.~Smolders, H.~Strathmann,
  \emph{Journal of membrane science} \textbf{1996}, \emph{113}, 361.

\bibitem{wang2013recent}
D.-M. Wang, J.-Y. Lai, \emph{Current Opinion in Chemical Engineering}
  \textbf{2013}, \emph{2}, 229.

\bibitem{haase2017multifunctional}
M.~F. Haase, H.~Jeon, N.~Hough, J.~H. Kim, K.~J. Stebe, D.~Lee, \emph{Nature
  communications} \textbf{2017}, \emph{8}, 1.

\bibitem{fernandez2021putting}
C.~Fern{\'a}ndez-Rico, T.~Sai, A.~Sicher, R.~W. Style, E.~R. Dufresne,
  \emph{JACS Au} \textbf{2021}, \emph{2}, 66.

\bibitem{forster2010biomimetic}
J.~D. Forster, H.~Noh, S.~F. Liew, V.~Saranathan, C.~F. Schreck, L.~Yang, J.-G.
  Park, R.~O. Prum, S.~G. Mochrie, C.~S. O'Hern, et~al., \emph{Advanced
  Materials} \textbf{2010}, \emph{22}, 2939.

\bibitem{lee2017chameleon}
G.~H. Lee, T.~M. Choi, B.~Kim, S.~H. Han, J.~M. Lee, S.-H. Kim, \emph{ACS nano}
  \textbf{2017}, \emph{11}, 11350.

\bibitem{shang2020photonic}
G.~Shang, M.~Eich, A.~Petrov, \emph{APL Photonics} \textbf{2020}, \emph{5},
  060901.

\bibitem{nagayama1996two}
K.~Nagayama, \emph{Colloids and Surfaces A: Physicochemical and Engineering
  Aspects} \textbf{1996}, \emph{109}, 363.

\bibitem{jiang1999single}
P.~Jiang, J.~Bertone, K.~S. Hwang, V.~Colvin, \emph{Chemistry of Materials}
  \textbf{1999}, \emph{11}, 2132.

\bibitem{matsubara2007thermally}
K.~Matsubara, M.~Watanabe, Y.~Takeoka, \emph{Angewandte Chemie International
  Edition} \textbf{2007}, \emph{46}, 1688.

\bibitem{edrington2001polymer}
A.~C. Edrington, A.~M. Urbas, P.~DeRege, C.~X. Chen, T.~M. Swager,
  N.~Hadjichristidis, M.~Xenidou, L.~J. Fetters, J.~D. Joannopoulos, Y.~Fink,
  et~al., \emph{Advanced Materials} \textbf{2001}, \emph{13}, 421.

\bibitem{valkama2004self}
S.~Valkama, H.~Kosonen, J.~Ruokolainen, T.~Haatainen, M.~Torkkeli, R.~Serimaa,
  G.~ten Brinke, O.~Ikkala, \emph{Nature materials} \textbf{2004}, \emph{3},
  872.

\bibitem{wu2018flexible}
C.-S. Wu, P.-Y. Tsai, T.-Y. Wang, E.-L. Lin, Y.-C. Huang, Y.-W. Chiang,
  \emph{Analytical chemistry} \textbf{2018}, \emph{90}, 4847.

\bibitem{patel2020tunable}
B.~B. Patel, D.~J. Walsh, D.~H. Kim, J.~Kwok, B.~Lee, D.~Guironnet, Y.~Diao,
  \emph{Science Advances} \textbf{2020}, \emph{6}, eaaz7202.

\bibitem{dufresne2009self}
E.~R. Dufresne, H.~Noh, V.~Saranathan, S.~G. Mochrie, H.~Cao, R.~O. Prum,
  \emph{Soft Matter} \textbf{2009}, \emph{5}, 1792.

\bibitem{saranathan2012structure}
V.~Saranathan, J.~D. Forster, H.~Noh, S.-F. Liew, S.~G. Mochrie, H.~Cao, E.~R.
  Dufresne, R.~O. Prum, \emph{Journal of The Royal Society Interface}
  \textbf{2012}, \emph{9}, 2563.

\bibitem{prum2006anatomy}
R.~O. Prum, \emph{Bird coloration, volume 1: mechanisms and measurements}
  \textbf{2006}, \emph{1}, 295.

\bibitem{saranathan2021evolution}
V.~Saranathan, S.~Narayanan, A.~Sandy, E.~R. Dufresne, R.~O. Prum,
  \emph{Proceedings of the National Academy of Sciences} \textbf{2021},
  \emph{118}, e2101357118.

\bibitem{radja2019pollen}
A.~Radja, E.~M. Horsley, M.~O. Lavrentovich, A.~M. Sweeney, \emph{Cell}
  \textbf{2019}, \emph{176}, 856.

\bibitem{sumper2002phase}
M.~Sumper, \emph{Science} \textbf{2002}, \emph{295}, 2430.

\bibitem{lenoci2008diatom}
L.~Lenoci, P.~J. Camp, \emph{Langmuir} \textbf{2008}, \emph{24}, 217.

\bibitem{feofilova2022geometrical}
M.~Feofilova, S.~Sch{\"u}epp, R.~Schmid, F.~Hacker, H.~T. Spanke, N.~Bain,
  K.~E. Jensen, E.~R. Dufresne, \emph{Proceedings of the National Academy of
  Sciences} \textbf{2022}, \emph{119}, e2201014119.

\bibitem{sicher2021structural}
A.~Sicher, R.~Ganz, A.~Menzel, D.~Messmer, G.~Panzarasa, M.~Feofilova, R.~O.
  Prum, R.~W. Style, V.~Saranathan, R.~M. Rossi, et~al., \emph{Soft Matter}
  \textbf{2021}, \emph{17}, 5772.

\bibitem{sheppard1979selection}
C.~S. Sheppard, V.~R. Kamath, \emph{Polymer Engineering \& Science}
  \textbf{1979}, \emph{19}, 597.

\bibitem{matyjaszewski2012atom}
K.~Matyjaszewski, \emph{Macromolecules} \textbf{2012}, \emph{45}, 4015.

\bibitem{bisht2001living}
H.~S. Bisht, A.~K. Chatterjee, \emph{Journal of Macromolecular Science, Part C:
  Polymer Reviews} \textbf{2001}, \emph{41}, 139.

\bibitem{matyjaszewski2018advanced}
K.~Matyjaszewski, \emph{Advanced Materials} \textbf{2018}, \emph{30}, 1706441.

\bibitem{goto2004kinetics}
A.~Goto, T.~Fukuda, \emph{Progress in Polymer Science} \textbf{2004},
  \emph{29}, 329.

\bibitem{matyjaszewski2001atom}
K.~Matyjaszewski, J.~Xia, \emph{Chemical reviews} \textbf{2001}, \emph{101},
  2921.

\bibitem{hill2015expanding}
M.~R. Hill, R.~N. Carmean, B.~S. Sumerlin, \emph{Macromolecules} \textbf{2015},
  \emph{48}, 5459.

\bibitem{pan2016photomediated}
X.~Pan, M.~A. Tasdelen, J.~Laun, T.~Junkers, Y.~Yagci, K.~Matyjaszewski,
  \emph{Progress in Polymer Science} \textbf{2016}, \emph{62}, 73.

\bibitem{rosen2009single}
B.~M. Rosen, V.~Percec, \emph{Chemical Reviews} \textbf{2009}, \emph{109},
  5069.

\bibitem{flory1942thermodynamics}
P.~J. Flory, \emph{The Journal of chemical physics} \textbf{1942}, \emph{10},
  51.

\bibitem{kwak2011arget}
Y.~Kwak, A.~J. Magenau, K.~Matyjaszewski, \emph{Macromolecules} \textbf{2011},
  \emph{44}, 811.

\bibitem{simakova2012aqueous}
A.~Simakova, S.~E. Averick, D.~Konkolewicz, K.~Matyjaszewski,
  \emph{Macromolecules} \textbf{2012}, \emph{45}, 6371.

\bibitem{whitfield2021low}
R.~Whitfield, K.~Parkatzidis, K.~G. Bradford, N.~P. Truong, D.~Konkolewicz,
  A.~Anastasaki, \emph{Macromolecules} \textbf{2021}, \emph{54}, 3075.

\bibitem{chan2008arget}
N.~Chan, M.~F. Cunningham, R.~A. Hutchinson, \emph{Macromolecular Chemistry and
  Physics} \textbf{2008}, \emph{209}, 1797.

\bibitem{mac2019application}
N.~Mac~Fhionnlaoich, R.~Qi, S.~Guldin, \emph{Langmuir} \textbf{2019},
  \emph{35}, 16605.

\bibitem{boddeker2022non}
T.~J. B{\"o}ddeker, K.~A. Rosowski, D.~Berchtold, L.~Emmanouilidis, Y.~Han,
  F.~H. Allain, R.~W. Style, L.~Pelkmans, E.~R. Dufresne, \emph{Nature physics}
  \textbf{2022}, \emph{18}, 571.

\bibitem{glatter1982small}
O.~Glatter, O.~Kratky, H.~Kratky, \emph{Small angle X-ray scattering}, Academic
  press \textbf{1982}.

\bibitem{perez2016structure}
M.~P{\'e}rez~M{\'e}ndez, D.~Rodr{\'\i}guez~Mart{\'\i}nez, J.~Fayos
  \textbf{2016}.

\end{thebibliography}

\newpage

\end{document}